\documentclass[prl,reprint,amsmath,amssymb,preprintnumbers,superscriptaddress,nofootinbib]{revtex4-1}
\usepackage[dvipdf,dvips,dvipdfmx]{graphicx}
\usepackage{amsmath}
\usepackage{amssymb}
\usepackage{float}
\usepackage{wrapfig}
\usepackage{bm}
\usepackage{color}
\usepackage{comment}

\newcommand{\red}[1]{{\color{black}#1}}

\newcommand{\al}[1]{\begin{align}#1\end{align}}

\newcommand{\paren}[1]{\left(#1\right)}

\newcommand{\sqbr}[1]{\left[#1\right]}

\newcommand{\vev}[1]{\left\langle#1\right\rangle}

\newcommand{\nn}{\nonumber\\}

\newcommand{\p}{\partial}
 
\begin{document}
\preprint{MAD-TH-18-01
}
\title{
Infinite Set of Soft Theorems in Gauge-Gravity Theories as Ward-Takahashi Identities
}  
\date{\today}

\author{
Yuta Hamada
}
\email{
yhamada@wisc.edu
}
\affiliation{
Department of Physics, University of Wisconsin-Madison, Madison, WI 53706, USA
}
\affiliation{
KEK Theory Center, IPNS, KEK, Tsukuba, Ibaraki 305-0801, Japan
}
\author{Gary Shiu}\email{shiu@physics.wisc.edu}
\affiliation{
Department of Physics, University of Wisconsin-Madison, Madison, WI 53706, USA
}




\begin{abstract}
We show that the soft photon, gluon and graviton theorems can be understood as the Ward-Takahashi identities of large gauge transformation, i.e., diffeomorphism that does not fall off at spatial infinity. We found infinitely many new identities which constrain the higher order soft behavior of the gauge bosons and gravitons in scattering amplitudes of gauge and gravity theories. Diagrammatic representations of these soft theorems are presented.
\end{abstract}
\maketitle

\normalsize

\section*{Introduction}
The last few decades have witnessed a remarkable synergy between particle physics and cosmology. While the objects of interest are vastly different, the tools to study them often share a great deal of similarity.  A recent example of this synergy is the notion of ``cosmological collider physics" \cite{Arkani-Hamed:2015bza}. 
Just as how we extract particles' interactions from scattering in colliders, we can extract the interactions governing the early universe from the correlations of primordial fluctuations that seed cosmic structure formation. Thus, $N$-point corrections ($N\geq3$) of the curvature perturbation 
(aka non-Gaussianities) play a similar role as scattering amplitudes in particle physics.

Another commonality between particle physics and cosmology is that they are both populated with a myriad of theories and the challenge is to zero-in on the right ones. In this regard, results based on symmetries are especially powerful as they allow us to test and discriminate broad classes of theories in a model-independent manner.
A prominent example are soft theorems which govern the behavior of scattering amplitudes or correlation functions when one or more of the momenta approach zero.
The subject of soft theorems has a long history in particle physics, dating back to the early studies of the soft pion theorem.
The equivalence between the soft photon theorem~\cite{Low:1958sn, GellMann:1954kc} and large gauge transformation has already been noted long ago by Ferrari and Picasso~\cite{Ferrari:1971at} (see also \cite{Ferrari:1970pe}), where the leading and subleading soft theorems were shown to follow from the Ward-Takahashi (WT) identity of linear transformation which survives after Lorenz gauge fixing.

Recently, there has been a revival of interest in soft theorems, due largely to the series of work by Strominger et al~\cite{Strominger:2013lka,Strominger:2013jfa,He:2014laa} (see Ref.~\cite{Strominger:2017zoo} for a review). Along these lines, new subsubleading tree level soft theorems were found \cite{Cachazo:2014fwa}. These results on higher order soft behaviors supplement earlier findings on 
 the leading and subleading soft photon and graviton theorems \cite{Weinberg:1965nx,Gross:1968in,Jackiw:1968zza}
 (analogous soft gluon theorem up to subleading order can be found in Refs.~\cite{Casali:2014xpa,Berends:1988zn,Mao:2017tey}).
Nonetheless, methods based on asymptotic charges (symmetry charges of large gauge transformations that survive at null infinity) have their limitations as they do not  seem to lead to soft theorems beyond the subsubleading order.

In the context of inflationary cosmology, similar soft theorems have been obtained. The first known example is Maldacena's 
consistency condition \cite{Maldacena:2002vr} which states that the three-point function of the curvature perturbation $\zeta$ (also known as the adiabatic mode \cite{Weinberg:2003sw}) in the squeezed  limit where one of the modes becomes soft, is determined by the scale transformation of the two-point function:
\begin{equation}
\lim_{\vec{q} \rightarrow 0} \frac{1}{P_{\zeta} (q)} \vev{ \zeta_{\vec{q}} \zeta_{\vec{k}_1} \zeta_{\vec{k}_2} }^{\prime}
= - \vec{k}_1 \cdot \frac{\partial}{\partial_{\vec{k}_1} } \vev{ \zeta_{\vec{k}_1} \zeta_{\vec{k}_2}}^{\prime}
\label{Maldacena}
\end{equation}
where $P_{\zeta} (q)$ denotes the power spectrum, and \red{$\vev{\dots}^{\prime}$} are correlators without the momentum-conserving $\delta$-function.
Subsequently, an infinite set of WT identities for the adiabatic mode have been derived in Ref.~\cite{Hinterbichler:2013dpa}, with the leading $q^0$ behavior of the soft limit recovering Eq.~(\ref{Maldacena}).
This infinite set of WT identities was also shown to be equivalent to the Slavnov-Taylor identity of spatial diffeomorphism~\cite{Berezhiani:2013ewa}, 
with the adiabatic mode argued to be related to the locality of the theory.\footnote{The method  of \cite{Hinterbichler:2013dpa} has been applied to flat spacetime in Ref.~\cite{Mirbabayi:2016xvc}, but only the leading soft theorem was derived there.}

The parallel between inflationary correlators and scattering amplitudes raises the following puzzle: can we obtain all the $q^n$ soft behavior of scattering amplitudes in gauge and gravity theories from WT identities of large gauge transformations?
In this Letter, we derive an infinite set of soft theorems for soft photon, gluon and graviton amplitudes\footnote{\red{Similar approach to the subleading soft graviton theorem can be found in Ref.~\cite{Campoleoni:2017mbt} though, unlike our present work, self-interactions of the graviton were neglected.}}
utilizing the WT identity of large gauge transformation/diffeomorphism.
As shown in Refs.~\cite{Broedel:2014fsa,Bern:2014vva}, on-shell gauge invariance determines i) the leading and subleading behavior of soft photons and soft gluons, and ii) the leading, subleading and subsubleading behavior of soft gravitons. 
As we shall see, our results are consistent with these previous works, but we further obtain constraints on the higher order $q^n$ terms coming from the infinitely many WT identities.
We focus on tree level processes in this Letter. The generalization to loop corrected amplitudes is left for future work.

Furthermore, we comment on the relation with the approach using asymptotic charges.
In Refs.~\cite{Conde:2016csj,Conde:2016rom}, it was argued that the $1/r$ expansion of large gauge transformations and supertranslation charges correspond to the (sub)subleading soft theorems of the photon/graviton. In this sense, our $x^n$ expansion seems to correspond to the $1/r$ expansion.
However, the method in Refs.~\cite{Conde:2016csj,Conde:2016rom} cannot determine the constant of integration completely, and higher order charges do not lead to soft theorems beyond the (sub)subleading order. 
As a different approach, it was shown in Refs.~\cite{Campiglia:2016hvg,Campiglia:2016jdj,Campiglia:2016efb} that
the subleading soft photon and subsubleading soft graviton theorems can be interpreted as asymptotic charge conservation associated with large gauge transformation or diffeomorphism. 

 It would be interesting to compare our method with the approach based on asymptotic charges.
Here, we note the merits of our approach, in relation to previous works.
First, it is easy to extend our analysis to any order in $\mathcal{O}(q^n)$, and indeed we found infinitely many new constraints. Second, it is similarly straightforward to treat massive particle as well as massless particle with our method. Third, the extension to the higher dimensional case is straightforward. Fourth, the effect of the higher dimensional operator can be easily included (see also Ref.~\cite{Elvang:2016qvq}). 
We emphasize that our method needs not  assume that the charged particle is massless or minimal coupled.
Our main results, Eqs.~\eqref{Eq:soft photon result}\eqref{Eq:soft gluon result}\eqref{Eq:soft graviton result}, just rely on symmetries, and so are applicable to theories with general masses and couplings.

\section*{WT identity for large gauge transformation and diffeomorphism}

Out starting point is the WT identity (see Refs.~\cite{Ferrari:1971at,Ezawa:1966zz} for the derivation) for the large gauge transformation\footnote{See also Ref.~\cite{Avery:2015rga} for the derivation of a WT identity of the correlation functions for residual gauge symmetry based on the path integral formalism.}, which is given by
\al{\label{Eq:WT identity}
&\lim_{R\to\infty}\langle 0| Q_{R,\alpha} E_1 B - B E_1 Q_{R,\alpha}|0\rangle
=
\lim_{R\to\infty}\langle 0| [Q_{R,\alpha}, B] |0\rangle,
\\&
Q_{R,\alpha}:=
\int d^3x\, f_R(\vec{x}) J_0. 
}
Here $f_R(\vec{x})=1$, $0$ for $|\vec{x}|<R$ and $|\vec{x}|>R$, respectively, $B$ is an arbitrary operator, and $E_1=\sum_\text{massless} |n\rangle \langle n|$ is the projection on the zero-mass one-particle states of the theory.
On the LHS, $R\to\infty$ corresponds to the soft limit of the gauge particle~\cite{Ferrari:1971at}, and the RHS is the expectation value of the transformation of $B$, and corresponds to the amplitude without the soft particle.

\begin{figure}
\begin{center}
\hfill
\includegraphics[width=.23\textwidth]{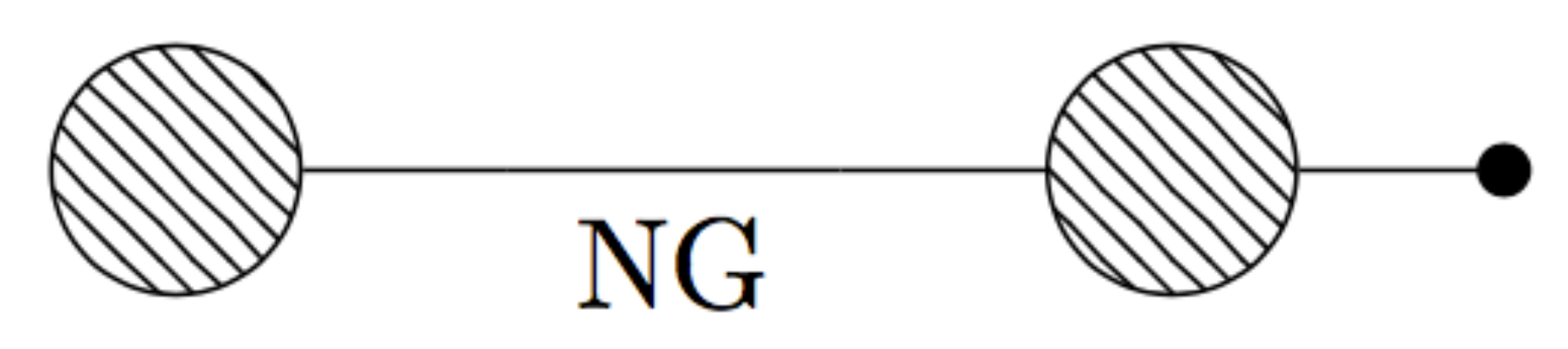}
\hfill
\includegraphics[width=.23\textwidth]{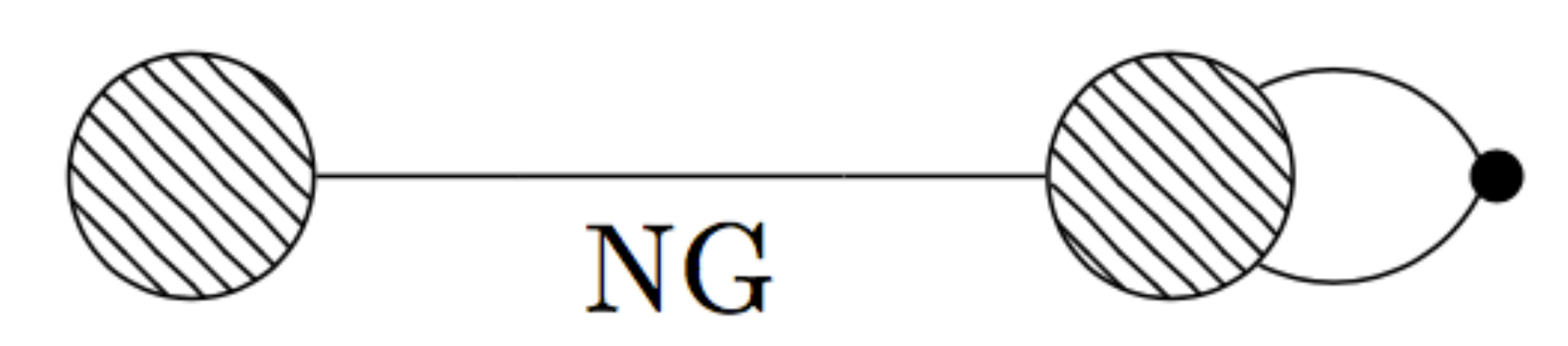}
\\
\hfill
\includegraphics[width=.23\textwidth]{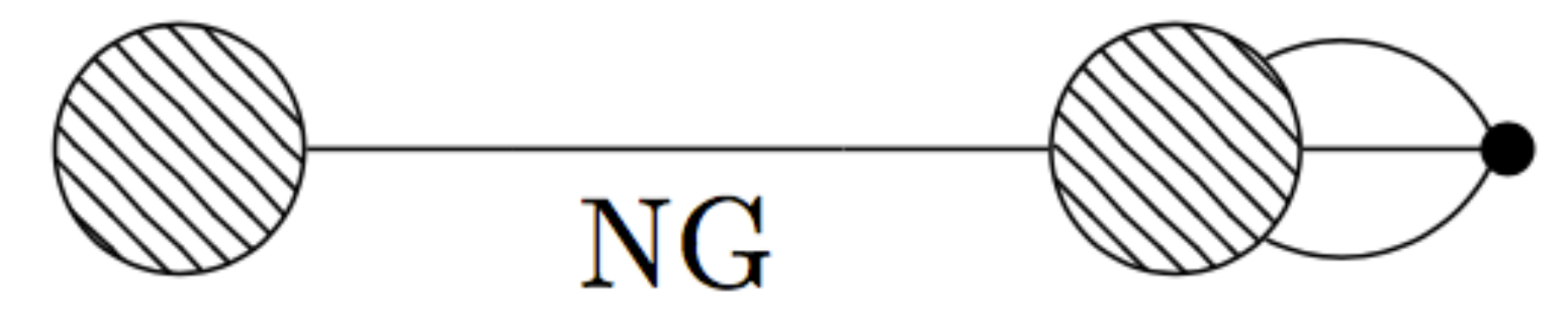}
\hfill\mbox{}
\end{center}
\caption{
The Feynman diagrams which contribute to the LHS of the WT identity. The black dot represents the insertion of the current $J_0$. There are infinitely many diagrams where $n$-legs attach with $J_0$, and we show the first three diagrams here. Since the intermediate state should be a Nambu-Goldstone (NG) one particle state, only the first diagram gives contribution at  tree level.
}
\label{Fig:WT_LHS}
\end{figure}

We can schematically express the transformation as $\delta\Phi_i=c_{0i}+c_{1ij}\Phi_j+c_{2ijk}\Phi_j\Phi_k+\cdots$. In general, the transformation of the gauge particle starts from the zeroth order while the that of matter particle starts from the linear order.
Since we are interested in the scattering amplitude, we perform the LSZ procedure for the $n$-point amplitude for both sides. Interestingly, at least at the tree level, only the linear transformation part survives on RHS.\footnote{See Ref.~\cite{Bianchi:2016viy} for the proof of Adler's zero in similar way.}
On the contrary, only the term in $J_0$ corresponding to the zeroth order transformation gives the contribution to the LHS at the tree level.
This is because the term in $J_0$ corresponding to the $n$-th order should contain $n+1$ number of the field, and can only be connected with the one-particle massless state at the loop level, see Fig.~\ref{Fig:WT_LHS}.
The detailed discussion of this point will be given in a subsequent paper.

\section*{Soft theorems from WT identities}
Applying Eq.~\eqref{Eq:WT identity}, we obtain an infinite set of WT identities. The point is that the we only need to consider the zeroth and linear order transformation to derive the tree level identity.\footnote{This situation is similar to Ref.~\cite{Hinterbichler:2013dpa}.}
We derive the soft photon, gluon, and graviton theorems
in theories consisting of the gauge particle and a (complex) charged scalar. 
Analogous soft theorems for theories with fermionic matter fields can be similarly obtained by our method outlined here.
\subsection*{Soft photon theorem}
We take the Lorenz gauge and the residual symmetry is $\p^2\chi=0$, where $\chi$ is the gauge transformation parameter, namely, 
\al{\label{Eq:photon residual symmetry}
&
\chi=\sum_{M=1}^\infty\eta_{i_1 i_2... i_M} x^{i_1} x^{i_2}\cdots x^{i_M},
&&
\eta_{ii i_3... i_n}=0,
}
where $\eta$ is totally symmetric and traceless. For simplicity, we assume that $\chi$ only depends on the spatial coordinate, which is sufficient to reproduce the known
soft theorems and derive new identities.\footnote{It might be possible to find new identities if we include the time coordinate. We leave this for future study.}

\begin{figure}
\begin{center}
\includegraphics[width=.35\textwidth]{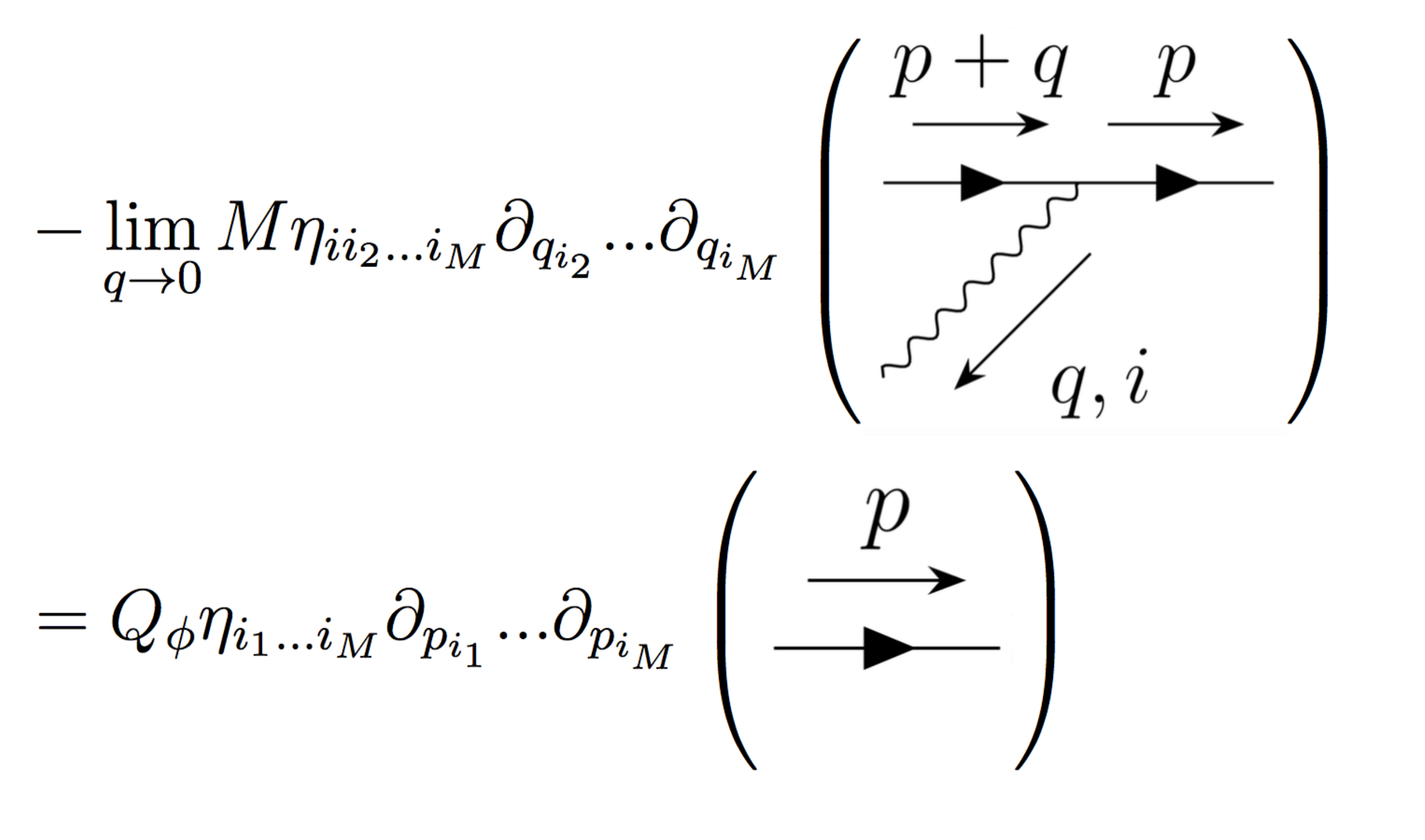}
\end{center}
\caption{
Diagrammatic representation of Eq.~\eqref{Eq:soft photon two-point}.
}
\label{Fig:2point_diagramatic_interpretation}
\end{figure}

\begin{figure}
\begin{center}
\includegraphics[width=.4\textwidth]{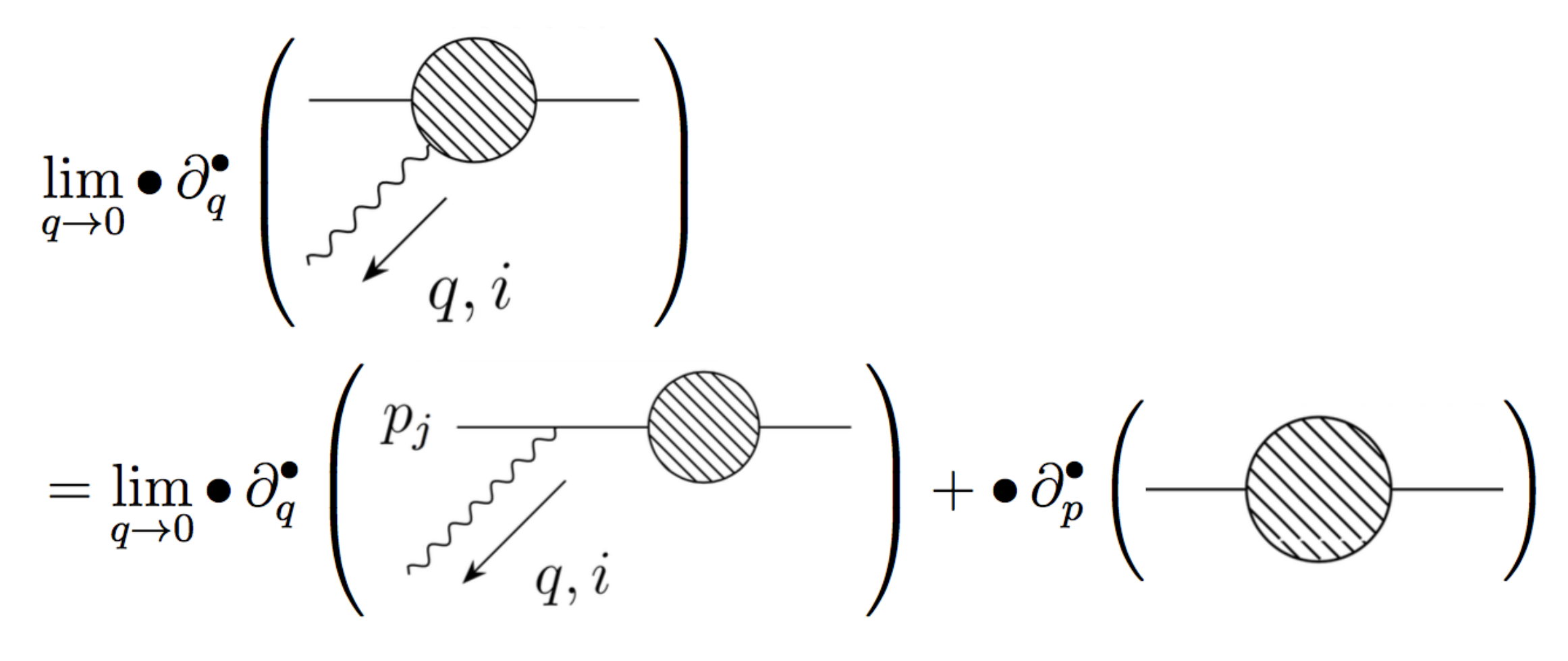}
\end{center}
\caption{
Schematic diagrammatic representation of the soft photon theorem, Eq.~\eqref{Eq:soft photon result}.
}
\label{Fig:Full_diagramatic_interpretation}
\end{figure}
Let us first apply the WT identity to the two-point functions:\footnote{See Ref.~\cite{Ferrari:1971at} for the derivation of $M=1$ case. We extend this result to general $M$.}
{\footnotesize
\al{
\p_i\chi|_{x=-i\p_q}S(p)\Gamma^i(-p-q,p,q)S(p+q)=\int d^4x\,e^{ipx}\vev{\delta\phi^*(x)\phi(0)}.
}}More explicitly, we obtain
{\footnotesize
\al{\label{Eq:soft photon two-point}
&-\lim_{q\to0}M\eta_{ii_2...i_M}\p_{q_{i_2}}...\p_{q_{i_M}}S(p)\Gamma^i(-p-q,p,q)S(p+q)
\nn&=Q_\phi\eta_{i_1...i_M}\p_{p_{i_1}}...\p_{p_{i_M}}S(p),
}}where $Q_\phi$ is the $U(1)$ charge of the matter, $S$ is the matter propagator, $\Gamma^\mu(p,q,r)$ is the three-point vertex with outgoing momenta $(q,p,r)$, and $\chi=\mathcal{O}(x^M)$. See Fig.~\ref{Fig:2point_diagramatic_interpretation} for the diagrammatic representation of Eq.~\eqref{Eq:soft photon two-point}.

Next, we consider the $(2m+n)$-point function, $B=T\paren{A^{\mu_1}(x_1)...A^{\mu_n}(x_n)\phi(y_1)...\phi(y_m)\phi^*(z_1)...\phi^*(z_m)}$, i.e., the time-ordered product of $m$-charged matter, $m$-charged anti-matter, and $n$-gauge bosons. The momenta of these particles are denoted by $(p_1,...,p_m,p'_1,...,p'_m,k_1,...k_n)$.
The WT identities give:\footnote{
As a convention, we take all external momenta to be outgoing direction.
}
{\footnotesize
\al{\label{Eq:soft photon result}
&\lim_{q\to0}\eta_{ii_2...i_M}\p_{q_2}...\p_{q_M}\mathcal{M}_{n+2m+1}^i\paren{q;p,p',k}
\nn&=
\lim_{q\to0}\sum_{j=1}^m
\eta_{ii_2...i_M}\p_{q_{i_2}}...\p_{q_{i_M}}
\nn& \times
 \bigg[ \Gamma^i(-p_j-q,p_j,q)S(p_j+q)\mathcal{M}_{n+2m}(p_j+q)  
\nn&
+\mathcal{M}_{n+2m}(p'_j+q)S(-p'_j-q)\Gamma^i(-p'_j-q,p'_j,q)\bigg]
\nn&+{Q_\phi\over M}\eta_{i_1...i_M}\sqbr{\p_{p_{i_1}}...\p_{p_{i_M}}\mathcal{M}_{n+2m} + (-1)^{M-1} (p\leftrightarrow p')},
}
}where $\mathcal{M}_{n+2m+1}, \mathcal{M}_{n+2m}$ are the amplitudes with and without the soft photon $(q,i)$; 
$q$, $i$ are the momentum and the polarization respectively.
The RHS of Eq.~\eqref{Eq:soft photon result} is derived by the following steps.
First we replace $\delta\phi$ and $\delta\phi^*$ on the right hand side
by
 the derivatives with respect to the external momentum $p, p'$ in the Fourier space. If the derivative acts on an external line, we can use Eq.~\eqref{Eq:soft photon two-point} to obtain the $q$-derivative. If the derivative acts on an internal line or a vertex, we can also obtain the $q$-derivative as\footnote{Since the momentum $q$ is introduced as in Eq.~\eqref{Eq:soft photon two-point}, the argument of  $\mathcal{M}_{n+2m}$ should become $p_j\to p_j+q$ in order to maintain momentum conservation (translation invariance) of the amplitude.} $\p_{p_j} \mathcal{M}_{n+2m}(p_j+q)=\p_q \mathcal{M}_{n+2m}(p_j+q)$. In this way, the $p$-derivatives become $q$-derivatives. On the other hand, the last term in Eq.~\eqref{Eq:soft photon result} corresponds to terms with derivatives of $\mathcal{M}_{n+2m}$. 
The diagrammatic representation of the identity is clear.
The amplitude $\mathcal{M}_{n+2m+1}^i$ consists of diagrams where the soft photon is attached to an external line, an internal line or a vertex. Thus, in principle, we have to consider all kinds of diagrams. However, the identity says that if the derivative $\eta_{ii_2...i_M}\p_{q_2}...\p_{q_M}$ is acting on  $\mathcal{M}_{n+2m+1}^i$, it can be determined by the diagram where the soft photon is attached only to an external line, and the diagram without the soft photon, see Fig.~\ref{Fig:Full_diagramatic_interpretation}. For $M = 1$, we can completely fix the amplitudes.
The $M=1$ case is known in Ref.~\cite{Ferrari:1971at} and is equivalent to the Low's subleading theorem~\cite{Low:1958sn}.
For higher order, we identified model independent part by using the projector.
These $M\geq2$ identities are new, and constrain the higher order soft photon amplitude.

\subsection*{Soft gluon theorem}
Again we take the Lorenz gauge. Contrary to the $U(1)$ gauge theory we cannot write down the exact $\chi^a$ which preserves the Lorenz gauge in a simple way because of the existence of hard gluons.
However, in principle, we can calculate the gauge parameter $\chi^{(n)a}$ which preserves the Lorenz gauge order by order of the gauge field. 
The zeroth and the first order solutions are sufficient for our purpose.
To the zeroth order of the gauge field, we have $\p^2\chi^{(0)a}=0$:
\al{&
\chi^{(0)a}=\sum_{n=1}^\infty\eta^a_{i_1 i_2... i_n} x^{i_1} x^{i_2}\cdots x^{i_n},
&&
\eta^a_{ii i_3... i_n}=0,
}
We concentrate on $t$-independent $\chi$.
To first order, we obtain
{\footnotesize
\al{
\p^2\chi^{(1)a}+g f^{abc} A_\mu^b \p^\mu \chi^{(0)c}=0,
\,
\to
\,
\chi^{(1)a}=-{1\over\p^2}g f^{abc} A_\nu^b \p^\nu \chi^{(0)c}.
}
}

Applying the WT identity to the two-point functions gives us:
{\footnotesize
\al{
&\p_i\chi|_{x=-i\p_q}D^{a_1a'_1,\mu_1\mu'_1}(p)\Gamma^{i\mu'_1\mu'_2,aa'_1a'_2}(q,p,-p-q)D^{a_2a'_2,\mu_2\mu'_2}(p+q)\nn&=\int d^4p\,e^{ipx}\vev{\delta A_{\mu_1}^{a_1}(x) A_{\mu_2}^{a_2}(0)}.
}}for the gluon two-point function, where $D$ is the gluon propagator, $\Gamma$ is the gluon three-point vertex, and
{\footnotesize
\al{
&\p_i\chi|_{x=-i\p_q}S^{i_1i'_1}(p)\Gamma_\text{matt}^{i,ai'_1j'_1}(q,p,-p-q)S^{j_1'j_1}(p+q)\nn&=\int d^4p\,e^{ipx}\vev{\delta\phi^*(x)\phi(0)}.
}}for the matter two-point function where $\Gamma_\text{mat}$ is the matter-gluon three-point vertex. Here $(\delta A_\mu)^a=\p_\mu\chi^a+g f^{abc}A_\mu^b\chi^c, \,\delta\phi=ig\chi^aT^a_\phi\phi$.

\begin{figure}
\begin{center}
\includegraphics[width=.45\textwidth]{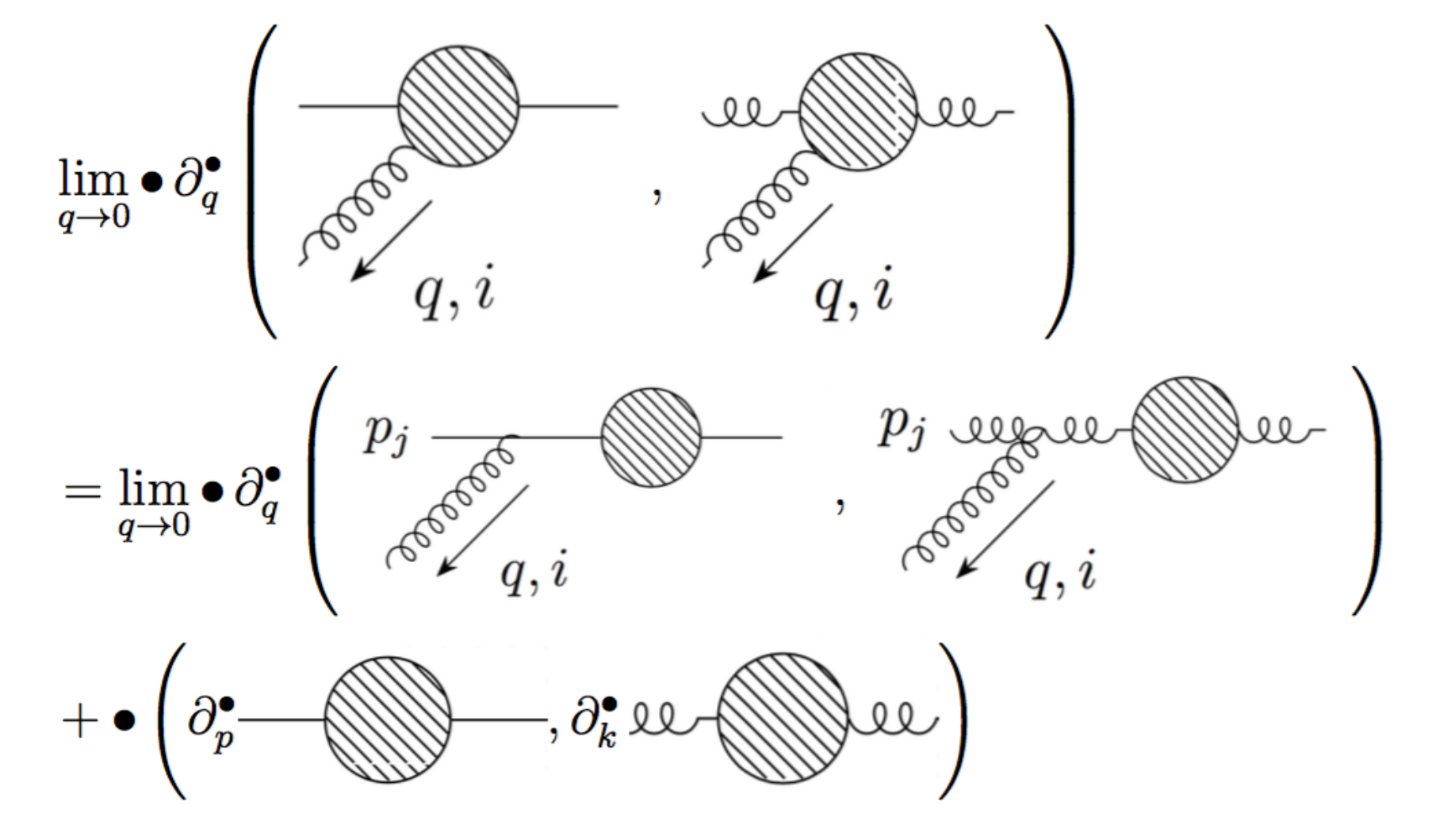}
\end{center}
\caption{
Schematic diagrammatic representation of the soft gluon theorem, Eq.~\eqref{Eq:soft gluon result}.
}
\label{Full_diagramatic_interpretation_gluon}
\end{figure}

By using the result of the two-point function, we can derive the identity for {\footnotesize$B=T\paren{A^{a_1\mu_1}_{x_1}...A^{a_n\mu_n}_{x_n}\phi^{i_1}_{y_1}...\phi^{i_m}_{y_m}\phi^{*j_1}_{z_1}...\phi^{*j_m}_{z_m}}$}:
{\footnotesize
\al{\label{Eq:soft gluon result}
&\lim_{q\to0}\eta^a_{i\alpha_2...\alpha_M}\p_{q_{\alpha_2}}...\p_{q_{\alpha_M}}\mathcal{M}_{n+2m+1}^{i,a}\paren{q;p,p',k}_{i_1...j_1...a_1...}
\nn&=
\lim_{q\to0}\sum_{l=1}^{m\,\text{or}\,n}\eta^a_{i\alpha_2...\alpha_M}\p_{q_{\alpha_2}}...\p_{q_{\alpha_M}}
\nn&\times\bigg[ {\Gamma^{i,ai_l i'_l}_\text{matt}(q,p_l,-p_l-q)S^{i'_l i''_l}(p_l+q)}\mathcal{M}_{n+2m}(p_l+q)_{i''_l}
\nn&+\mathcal{M}_{n+2m}(p'_l+q)_{j''_l}S^{j''_l j'_l}(-p'_l-q)\Gamma^{i,aj_l j'_l}_\text{matt}(q,p'_l,-p'_l-q)
\nn&
+ \Gamma^{i\mu_l\mu'_l,aa_la'_l}(q,k_l,-k_l-q)D^{a'_la''_l,\mu'_l\mu''_l}(k_l+q)\mathcal{M}_{n+2m}(k_l+q)_{a''_l}
\bigg]
\nn&+{g\over M}\eta_{i_1...i_M}\bigg[\p_{p_{i_1}}...\p_{p_{i_M}}(T^a_\phi)_{i_l i_l'}\mathcal{M}_{n+2m}(p_i,i_l') + (-1)^{M-1} (p\leftrightarrow p')
\nn&+\p_{k_{l,i_1}}...\p_{k_{l,i_M}}(T^a_\text{Ad})_{a_l a_l'}\mathcal{M}_{n+2m}(\mu_l,k_l,a_l')
\nn&-{Mk_{\mu_l}\over |\vec{k}|^2}\p_{k_{l,i_2}}...\p_{k_{l,i_M}}(T^a_\text{Ad})_{a_l a_l'}\mathcal{M}_{n+2m}(i_1,k_l,a_l')\bigg]
,
}}where ${\mathcal{M}_{n+2m+1}^{i,a}}(q;p,p',k)_{i_1...j_1...a_1...}$ is the amplitude with the soft gluon $(q, a)$.
The momenta and color indices of the hard charged scalars, anti-scalars, and gluons are
$(p_l, i_l)$, $(p'_l, j_l)$, and $(k_l,a_l)$ respectively.
$\mathcal{M}_{n+2m}$ is the amplitude without the soft gluon.
As in the photon case, we have checked that the $M=1$ case corresponds to the leading and subleading soft gluon theorems. The $M\geq2$ identities constrain the higher order terms. The interpretation of the identity is the same as in QED, see Fig.~\ref{Full_diagramatic_interpretation_gluon}.

\subsection*{Soft graviton theorem}
We take the following gauge,
\al{
ds^2=-dt^2+\paren{e^{\gamma}}_{ij}dx_idx_j,
}
where $\gamma_{ij}$ is transverse and traceless: $\p_i\gamma_{ij}=\gamma_{ii}=0$.
As in QCD, the hard graviton mode prevents us from finding an explicit solution of the gauge preserving $\xi_i$. We can write $\xi^{(0)}_i$ and the zeroth order transformation as
\al{&
\xi^{(0)}_i=\eta_{ii_1...i_n}x^{i_1}...\,x^{i_n},
&&
\delta \gamma_{ij}\sim\p_i\xi^{(0)}_j+\p_j\xi^{(0)}_i,
}
where $\eta_{ii_1...i_n}$ is traceless in the $(n+1)$ indices, and symmetric in the latter $n$ indices.
In the language of Ref.~\cite{Hinterbichler:2013dpa}, this is the ``tensor symmetry" which prevents the adiabatic mode from receiving time dependent corrections, see also Ref.~\cite{Mirbabayi:2016xvc}.
In principle, we can write down the linear transformation of $\gamma_{ij}$, but it is complicated, and the explicit form is not important.

As in other cases, we start from the application of the WT identity to the two-point amplitude.
It is found that
{\footnotesize
\al{
&\lim_{q\to0} (\p_i\xi^{(0)}_j+\p_j\xi^{(0)}_i)|_{x=-i\p_q} D^{(i_1j_1,i'_1j'_1)}(p)\Gamma^{(ij,i'_1j'_1,i'_2j'_2)}(q,p,-p-q)
\nn&\times D^{(i_2j_2,i'_2j'_2)}(p+q)
=
\int d^4p\,e^{ipx}\vev{\delta\gamma_{i_1j_1}(x)\gamma_{i_2j_2}(0)},
}}for the graviton and
{\footnotesize
\al{
&\lim_{q\to0} (\p_i\xi^{(0)}_j+\p_j\xi^{(0)}_i)|_{x=-i\p_q} S(p)\Gamma^{ij}(q,p,-p-q)S(p+q)\nn&
=\int d^4p\,e^{ipx}\vev{\delta\phi^*(x)\phi(0)},
}}for  the scalar. Here $\delta\phi=-\xi^i\p_i\phi$.

\begin{figure}
\begin{center}
\includegraphics[width=.45\textwidth]{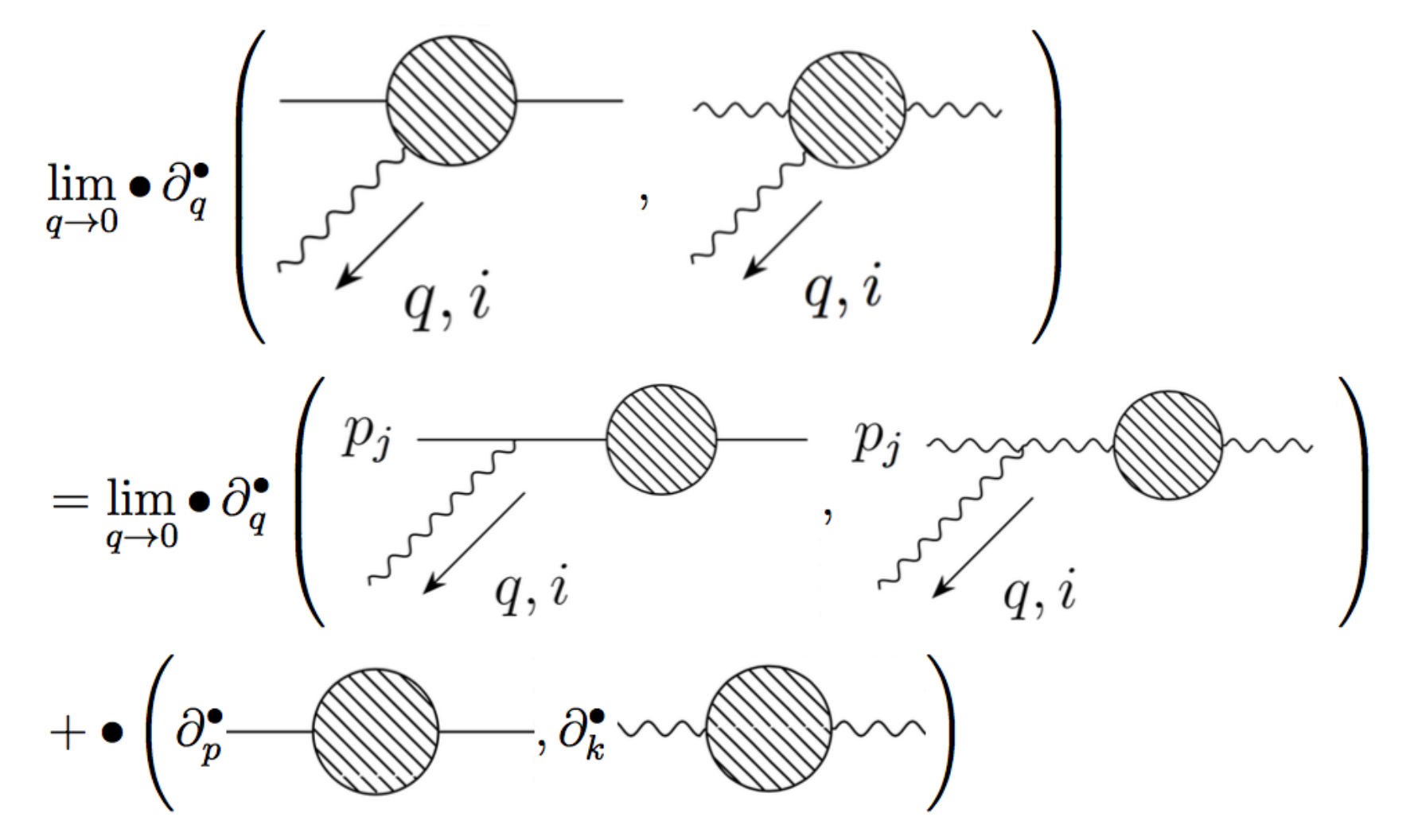}
\end{center}
\caption{
Schematic diagrammatic representation of the soft graviton theorem, Eq.~\eqref{Eq:soft graviton result}. The wavy line represents the graviton.
}
\label{Full_diagramatic_interpretation_graviton}
\end{figure}

Let us move on to the $(n+2m)$-point amplitude. 
By using the identity for the two-point function, we arrive at the following identity:
{\footnotesize
\al{\label{Eq:soft graviton result}
&\lim_{q\to0}\paren{\eta_{ij\alpha_2...\alpha_M}+\eta_{ji\alpha_2...\alpha_M}}\p_{q_{\alpha_2}}...\p_{q_{\alpha_M}}\mathcal{M}_{n+2m+1}^{ij}
\nn&=
\lim_{q\to0}\sum_{l}\paren{\eta_{ij\alpha_2...\alpha_M}+\eta_{ji\alpha_2...\alpha_M}}\p_{q_{\alpha_2}}...\p_{q_{\alpha_M}}
\nn&\times
\bigg[
\Gamma^{ij}_\text{matt}(q,p_l,-p_l-q)S(p_l+q)\mathcal{M}_{n+2m}(p_l+q)
\nn&+\mathcal{M}_{n+2m}(p'_l+q)S(-p'_l-q)\Gamma^{ij}_\text{matt}(q,p'_l,-p'_l-q)
\nn&
+\Gamma^{(ij,i'_lj'_l,i_lj_l)}(q,k_l,-k_l-q)D^{(i'_lj'_l,i''_lj''_l)}(k_l+q)\mathcal{M}_{n+2m}(k_l+q)_{i''_lj''_l}
\bigg]
\nn&+(\text{All derivatives act on $M_{n+2m}$}).
}}
Here $\mathcal{M}_{n+2m}(p,p',k)_{(i_1j_1)...(i_nj_n)}$ is the $n+2m$-point amplitude and $(i_1j_1)...(i_nj_n)$ is the polarization of each graviton. The diagrammatic representation of the identity is the same as before, see Fig.~\ref{Full_diagramatic_interpretation_graviton}.
As in the soft photon and gluon theorems, the last term corresponds to the case where all  the derivates act on internal lines or vertices. 
The detailed form may not be important, but the important point is that this term can be expressed in terms of the amplitude without the soft graviton.

We have checked that the $M=1$ identity corresponds to the leading and subleading graviton theorems, and the $M=2$ identity is the subsubleading soft graviton theorem. 
The $M\geq3$  identities constrain the higher order soft graviton amplitude.

\section*{Discussion}
In closing, we outline several interesting future directions motivated by the present work.
As shown in the recent body of works (e.g. Refs.~\cite{Strominger:2013lka,Strominger:2013jfa,He:2014laa,Seraj:2016jxi,Hamada:2017gdg,Hamada:2017uot,Hamada:2017atr,Compere:2017wrj,Hamada:2017bgi}, and reviewed in Ref.~\cite{Strominger:2017zoo}), there is an intriguing relation between
the asymptotic symmetry at null infinity and the memory effect. A natural question is whether there is a similar relation between the large gauge transformation/diffeomorphism at spatial infinity and the memory effect.

In addition to soft theorems for photons, gluons, and gravitons, it would also be interesting to apply our method to studying the soft behavior of the Nambu-Goldstone mode. (for recent work on the single soft scalar/pion theorem, see Refs.~\cite{Campiglia:2017dpg,Hamada:2017atr,Low:2017mlh,Campiglia:2017xkp}).

Our result may also have implications for elucidating the symmetry constraints of quantum gravity.
In addition to the first few leading soft theorems that were previously known in the literature,
the identities for the higher order soft behavior that we found should further constrain the tree-level scattering amplitudes in 
any quantum theory of gravity.
It would be interesting to pursue this possibility, in conjunction with the study of soft theorems in string theory \cite{Sen:2017xjn}.
Another interesting direction is the generalization to the higher spin case~\cite{Campoleoni:2017mbt}. 

In this Letter, we focus on tree level processes.  It is known that soft theorems get corrections from the infrared singularity at the loop level except for the leading soft theorem~\cite{Bern:2014oka,He:2014bga}. Extending our method to the loop level
may shed some interesting light on these corrections, see Ref.~\cite{Guerrieri:2017ujb} for a related discussion.

\subsection*{Acknowledgement}
We used TikZ-Feynman~\cite{Ellis:2016jkw} to draw the Feynman diagram.
This work is supported in part by the Grant-in-Aid for Japan Society for the JSPS Fellows No.16J06151 (YH), the DOE grant DE-SC0017647 (GS) and the Kellett Award of the University of Wisconsin (GS).

\bibliographystyle{Letter}
\bibliography{Bibliography}


\end{document}